\documentclass[11pt]{article}
\newfont{\larom}{cmbx9 scaled\magstep3}
\newfont{\bsan}{cmssbx10}
\newcommand{\be}{\begin{equation}}
\newcommand{\ee}{\end{equation}}
\newcommand{\lb}{\label}
\newcommand{\apj}{{\it Astrophys.\ J.}}
\newcommand{\physrep}{{\it Phys.\ Rep.}}
\newcommand{\araa}{{\it Annual Rev.\ Astron.\ Astrophys.}}
\newcommand{\nat}{{\it Nature}}

\newcommand{\sci}{{\it Science}}
\newcommand{\aea}{{\it Astron.\ Astrophys.}} 
\newcommand{\mnras}{{\it Monthly Not.\ Royal Astron.\ Soc.}} 
\newcommand{\grg}{{\it Gen.\ Rel.\ Grav.}} 
 
\begin{document}
\begin{center}
  {\larom The Apparent Fractal Conjecture:\\Scaling
  Features in Standard Cosmologies}\\
  \vspace{10mm}
  {\Large Marcelo B.\ Ribeiro}\\
  \vspace{6mm}
  Physics Institute, University of Brazil - UFRJ, CxP 68532,
  Rio de Janeiro, RJ 21945-970, Brazil; E-mail: mbr@if.ufrj.br\\
  \vspace{10mm}
  {\bf ABSTRACT}
\end{center}
\begin{quotation}
  \small \it
  \noindent This paper presents an analysis of the smoothness problem in
  cosmology by focussing on the ambiguities originated in the simplifying
  hypotheses aimed at observationally verifying if the large-scale
  distribution of galaxies is homogeneous, and conjecturing that this
  distribution should follow a fractal pattern, in the sense of having a
  power-law type average density profile, in perturbed standard
  cosmologies. This is due to a geometrical effect, appearing when
  certain types of average densities are calculated along the past light
  cone. The paper starts by reviewing the argument concerning the
  possibility that the galaxy distribution follows such a scale invariant
  pattern, and the premises behind the assumption that the spatial
  homogeneity of standard cosmology can be observable. Next, it is argued
  that in order to discuss observable homogeneity one needs to make a
  clear distinction between local and average relativistic densities, and
  showing how the different distance definitions strongly affect them,
  leading the various average densities to display asymptotically opposite
  behaviours. Then the paper revisits Ribeiro's (1995) results, showing
  that in a fully relativistic treatment some observational average
  densities of the flat Friedmann model are not well defined at
  $z \sim 0.1$, implying that at this range average densities
  behave in a fundamentally different manner as compared to the linearity
  of the Hubble law, well valid for $z<1$. This conclusion brings into
  question the widespread assumption that relativistic corrections can
  always be neglected at low $z$. It is also shown how some key features
  of fractal cosmologies can be found in the Friedmann models. In view of
  those findings, it is suggested that the so-called contradiction between
  the cosmological principle, and the galaxy distribution forming an
  unlimited fractal structure, may not exist.
\end{quotation}
\begin{flushleft}
  PACS numbers:
        98.80.-k 
  \ \ \ 98.65.Dx 
  \ \ \ 98.80.Es 
  \ \ \ 05.45.Df 
\end{flushleft}
\begin{flushleft}
  KEY words: {\it cosmology: theory and observations; large-scale structure
              of the Universe:\\ \hspace*{2.1cm} fractals} 
\end{flushleft}
\newpage
\newpage
\begin{flushright}
      {\it ``The difficulty lies, not in the new ideas, but in escaping
      from\\the old ones, which ramify ($\ldots$) into every corner of our
      minds.''}\\John Maynard Keynes (1936)
\end{flushright}
\section{Introduction}\label{intro}
Cosmology, like astronomy, often needs to rely upon some transitory and
simplifying assumptions in order to be able to compare theory with
observations. As those assumptions, usually arising from technological
constraints, are aimed at bringing a difficult problem into a workable,
possibly falsifiable, level, the conclusions reached through them need,
therefore, to be revised from time to time. However, it is easy to see
{\it a posteriori} when a certain hypothesis has aged, but not so much
so at the very moment when certain simplifying ideas need revision, or
even abandonment. The reasons for that are many, but they often come
about when technical means for gathering scientific data evolves, bringing
new data which may be orders of magnitude more accurate than previously
available, and, at the same time, the theoretical impact of such new
and more precise data on those simplifying assumptions goes unnoticed.
Besides, the theoretical implications of such revisions tend to be
resisted, which in turn generates controversy, inasmuch as they may well
lead to thorny questions which, quite often, are only reluctantly asked.

The issue of the scale where the matter distribution in the Universe
would become {\it observationally} smooth involves that kind of
simplifying assumptions. At the heart of this issue lies the problem of
how one can {\it observationally} characterize cosmological density, and,
for that purpose it is usually assumed that relativistic corrections can
be neglected in cosmology at redshift ranges where distance and redshift
follow a linear relation, {\it i.e.}, the Hubble law.\footnote{ \ {From}
now on I shall call by ``small redshifts'' the scales where $z < 0.1$,
by ``moderate redshifts'' when we have $0.1 \le z < 1$, and by ``large
redshifts'' the scales where $z \ge 1$. The linearity of the
redshift-distance relation is known to be valid at small and moderate
redshift ranges (Sandage 1995, p.\ 91).} This can be thought of as being
the cosmological Newtonian approximation, since the usual interpretation
is that Newtonian cosmology represents a small and local piece of the
Universe (see, {\it e.g.}, Harrison 2000, p.\ 332), where Newtonian
mechanics was long ago found to lead to the same dynamical equations as
given by general relativity (Milne 1934, McCrea and Milne 1934; good
reviews on these results, and their implications, can be found in Bondi
1960, Sciama 1993, and Harrison 2000). This assumption as applied to
cosmology means that flat and Euclidean geometry is valid in this local
observable region, with relativistically derived expressions becoming
unnecessary in observational cosmology (Peebles 1980, p.\ 143). Then, the
reasoning goes, as Newtonian and Friedmann cosmologies have homogeneous
densities at the same epochs, or, stating the same in relativistic
terminology, they have homogeneous spatial sections at constant time
coordinates, if we take these models as our best physical representations
of the Universe, their spatial homogeneity should be observed up to at
least moderate redshift ranges. So, sources up to $z \approx  1$ are
still assumed to lay within our local and Newtonian piece of the Universe.
Those are such standard and widespread assumptions of observational
cosmology that they are hardly stated openly, being almost always
assumed implicitly.

Therefore, the possible observational smoothness of the Universe in fact
relies on two inter-dependent and simplifying hypotheses of observational
cosmology: (1) relativistic corrections may be safely disregarded in
dealing with astronomical data of cosmological importance up to moderate
redshift ranges, as at those limits we are supposed to be still probing
within the local Newtonian piece of the Universe, and (2) an uniform
distribution of mass can, in fact ought to, be inferred from astronomical
data gathered at this local Newtonian universe.\footnote{ \ Since the
observed Universe is filled with stars, galaxies, etc, the second
hypothesis above implies that this observed lumpiness of the Universe
must originate in very local disturbances from uniformity, a phenomenon
statistically similar to white noise. The concept of a {\it correlation
length} was introduced with the aim of finding the maximum range of
this disturbance before the uniform distribution is reached (Peebles
1993). It is therefore obvious that these two additional hypotheses,
that is, lumpiness identified with white noise and the correlation 
statistics, are a corollary of the second simplifying hypothesis above,
and, thus, they cannot survive independently from it.}

However, for theoretical and practical purposes, there are at least two
ways of finding the range of this Newtonian approximation. The most
widely used is to take small speeds as meaning this approximation, which
implies that when galaxy recession speeds are small, as compared to light
speed, Newtonian mechanics is valid (Callan, Dicke and Peebles 1965).
Less well-known, but equally important, is the criterion implicitly
advanced by Bondi (1960), which states that it is light dynamical
behaviour that determines this range. In other words, as opposed to
relativity, Newtonian mechanics does not produce a dynamical theory for
light and, therefore, there will be cosmological scales sufficiently
large such that light can no longer be considered as being instantaneously
transmitted from source to observer.

The important point here is that contrary to what may be initially
thought, the practical implementation of these two criteria does not
always lead to the same results. The first criterion relies upon the
Hubble law being approximately written as a velocity-distance law
(Harrison 1993), while Bondi's criterion means solving the null geodesic
in a fully relativistic model, obtaining expressions for the observational
quantities along the observer's past null cone, and comparing those
observables with the Newtonian predictions. However, as I shall show
below, in Bondi's criterion the non-linearity of general relativity will
mean that different observational quantities will have Newtonian
approximations at different ranges. Therefore, we must see those two
criteria as complementing each other, and this implies that the range of
the Newtonian approximation, and, as a consequence, the limit up to where
we can dismiss relativistic corrections in cosmology, will depend upon the
observational quantities being dealt with. In other words, those limits
will depend on the specific problem under analysis. 

In this paper I intend to discuss the problem of the observational
smoothness of the Universe, and the possibility that the large-scale
mass distribution may follow a scale invariant, or fractal, pattern,
in the light of Bondi's criterion as outlined above. For this aim it is
mandatory to start by discussing the theoretical problems relevant for
the observational characterization of density in the cosmology, namely
local versus average density, and distance definitions. For simplicity,
I will use the Einstein-de Sitter model, but the analysis and most results
are also valid for open and close Friedmann models. I will show that once
this method is consistently and systematically applied, the two basic
assumptions of observational cosmology relevant for the smoothness
problem of the Universe, as described above, breakdown at small
redshift ranges. Then we will be able to obtain some results long ago
hypothesized for a hierarchical, or fractal, universe, without any need to
drop out either the cosmological principle or the usual meaning of
cosmological parameters like the Hubble constant, or even the cosmic
microwave background radiation (CMBR) isotropy.\footnote{ \ In fact, the
approach presented here avoids a collision between the cosmological
principle, and CMBR near isotropy, with the fractal universe model
advanced by Pietronero and collaborators, as those issues become then
unrelated.} I will also show that if we do not address the cosmological
smoothness problem via a fully relativistic perspective of cosmological
density, we may end up with some misleading conclusions drawn from the
initial, but no longer valid, assumptions, which in turn will inevitably
lead to false problems. Finally, concluding my analysis, I will discuss
how those findings naturally lead to the conjecture that the observed
fractality of the distribution of galaxies, defined here as being a
system characterized by a power-law type average density profile which
decays linearly at increasing distances, should be an observational
effect of geometrical nature, arising in perturbed Friedmann models.

Throughout this paper I shall avoid strict astronomical issues, as well
as discussing the aspects of the distribution of galaxies correlation
statistics, which has been at the centre of the debate, as applied to
relativistic models, inasmuch as such a discussion can be found
elsewhere (Ribeiro 1995). Therefore, here I shall focus on answering the
questions of whether or not the standard cosmology really implies a
perfectly well defined {\it observable} mean density, and also if it
completely rules out an unlimited fractal pattern. Davis (1997) pointed
out that these questions revolve around the concept of mean density of
the Universe. However, I claim here that in fact these questions revolve
around the concept of {\it observational} mean density of the Universe.

The plan of the paper is as follows. In \S \ref{debate}, after briefly
reviewing the proposal that large-scale distribution of matter follows a
hierarchical, or fractal, pattern, as well as the orthodox claim of why
this is untenable, I will discuss how an observed fractality needs an average
density definition, and that differentiating local versus average density
is essential to discuss the smoothness problem in cosmology. This point
was already present in Wertz (1970, 1971) earlier contribution to the
subject.

Section \S \ref{densities} deals with the issue of how we can
define and use those two densities in a relativistic framework. I will
show that Bonnor's (1972) first attempt to discuss this problem within a
relativistic context, while conceptually precise, fell short of
making strict use of Bondi's criterion due to analytical difficulties,
and that led him to use an inappropriate distance definition. The point to be
made here is that behind an observational definition of density lies an
unavoidable choice to be made among the various cosmological distances.
This is an essential point in order to put the discussion about the universal
smoothness problem into a solid relativistic footing. \S \ref{distances}
deals with this problem, proposing some criteria to be used for finding
the appropriate distance definition for the problem under consideration.
Then \S \ref{both} shows unequivocally that the Friedmann cosmology does
not imply a perfectly well defined apparent mean density in all
scales,~\footnote{ \ In this paper the word apparent means the same as
observational, in relativistic terms. In other words, all apparent results
imply that they were derived along the null cone, as prescribed by Bondi's
method.} but also that this observational average density already becomes
not well defined at small scales ($z < 0.1$), simply because when one
attempts to observationally characterize the {\it geometrical} constant
local density of the Friedmann models at moderate redshifts by means of
Bondi's criterion, the high non-linearity of Einstein's field equations,
together with the fact that volumes increase three times faster than
distances, will amplify very small differences in some observables which,
in turn, lead to dramatic differences in the values of the observational
mean density even at small $z$. Therefore, in \S \ref{both} it becomes
clear that in order to properly characterize observational density we need
to depart from the observational relations usually found in the
literature, since those are only valid at very small redshifts. In other
words, we need to derive new observational relations capable of taking
into account the lookback time. This section also revisits some of
Ribeiro's (1995) results, showing that the linearity of the
distance-redshift relation does not help in the context of observational
characterization of cosmological density as this relation remains well
approximated by a linear relation throughout moderate redshift ranges,
without any change in the value of $H_0$. Thus, it cannot be used as a
test for possible dismissal of large-scale hierarchical (fractal)
clustering, as has been done in the past (Sandage, Tammann and Hardy
1972; Sandage and Tammann 1975).  Finally, \S \ref{conjecture} collects
all those results by proposing that the observed fractal structure should
arise in perturbed Friedmann cosmologies. The paper finishes with a
conclusion where I argue that the near CMBR isotropy brings no
difficulties to the scenario outlined here.

Some terms used in this paper are applied elsewhere with somewhat
different meanings. Therefore, to avoid confusion, I shall define them
immediately. Here {\it fractality} refers to the property shown by the
observed large-scale distribution of galaxies of having an average
density power-law type decay at increasing distances. So, in this paper,
and all others where I have so far dealt with this issue, fractality
means in fact {\it observational fractality}, in the astronomical sense,
and only resembles non-analytical fractal sets in the sense that if we
define a smooth-out average density on those sets, the properties of
this average density are similar to what is found in observational
cosmology data. In other words, they are both of power-law type ones.
So, fractality has here an operative definition which allows us to
talk about fractality, or fractal properties, in completely smooth
relativistic cosmological models, where the cosmological fluid
approximation is assumed.

By {\it observational smoothness}, or {\it observational homogeneity}
of the Universe, I mean the possibility of inferring from observational
cosmology data that the large-scale distribution of galaxies has
constant average density. This is {\it the smoothness problem in
cosmology}. Therefore, there must be a clear distinction between
observational homogeneity and {\it spatial homogeneity}. The latter is
a built-in geometrical feature defined in the standard cosmologies, while
the former is the possible direct observation of the latter.

The main theses presented in this paper were briefly outlined in
Ribeiro (2001). Here, however, they are introduced in a very different
context, where I also provide a greatly expanded discussion, together
with many additional quantitative details, new results, thoughts, and
conclusions. This paper is, therefore, a follow-up to Ribeiro (2001). 

\section{The Hierarchical (Fractal) Clumping of Matter}\label{debate}

The proposal that the large-scale distribution of matter in the Universe
should follow a hierarchical, or fractal pattern is by no means new,
dating back to almost a century ago. The first proposal was made even
before relativistic cosmology itself was born in 1917. The initial
suggestion that the Universe could be constructed in a hierarchical
manner dates back to the very beginnings of cosmology (Fournier D'Albe
1907; Charlier 1908, 1922), with contributions made even by Einstein
(Amoroso Costa 1929; Wertz 1970; Mandelbrot 1983; Ribeiro 1994;
Ribeiro and Miguelote 1998). Since then it has kept reappearing in
the literature, being ressurected by someone who questioned the
accepted wisdom. So, the survivability of this concept is a fact
which should call the attention of a future historian of science.
Besides, while fractals were easily accepted in many areas of physics
as bringing new and useful tools and concepts, it is amazing to
witness the stiff resistance so many researches have been waging
against the introduction of any kind of fractal ideas in cosmology,
a fact worthy of attention (Oldershaw 1997; Ribeiro and Videira 1998;
Disney 2000).

In any case, what we are witnessing now is only the latest chapter of
a century old debate, which is now focused on the statistical methods
used by cosmologists to study data on galaxy clustering, and whether or
not the large-scale galaxy distribution follows a scaling pattern, in
the sense of having a power-law average density profile. The previous
chapter was between de Vaucouleurs (1970ab) and Wertz (1970, 1971; see
also de Vaucouleurs and Wertz 1971) on one side, and Sandage, Tammann,
and Hardy (1972) on the other side, and was mainly focused on
measurements of galaxy velocity fields and deviations from uniform
expansion, a topic which has also resurfaced in the recent debate
(Coles 1998).

\subsection{The Fractal Debate}

The latest round surrounding the smoothness problem in cosmology has
become known as {\it The Fractal Debate}. The controversy started with
Pietronero's (1987) claims that the usual correlation statistics
employed in the characterization of the distribution of galaxies cannot
be applied to this distribution, and that a novel statistical technique
proposed by him is capable of testing, rather than assuming, whether or
not the galaxy distribution is uniform.\footnote{ \ An earlier edition
of Mandelbrot's book on fractals led Peebles to discuss, and dismiss,
the possibility of an unlimited fractal pattern in the distribution of
galaxies, as proposed by Mandelbrot (see Peebles 1980, pp.\ 243-249,
and references therein). The reasoning behind his dismissal was, however,
again based on neglecting relativistic effects at small redshifts
(ibid., p.\ 245). Nevertheless, he kept open the possibility that other
fractal sets could provide a better modelling (ibid., p.\ 249).
This preview of The Fractal Debate seems to had attracted little
attention, and interest, in the astronomical/cosmological community.}
The main results reached by this side of the debate are the absence of
any sign of homogenization of the distribution of galaxies up to the
limits of current observations, denying, thus, any usefulness to a
correlation length concept (see \S \ref{intro} above), and that this
distribution is well described as forming a single fractal structure,
with dimension $D \approx 2$ (Coleman and Pietronero 1992; Pietronero
et al.\ 1997; Ribeiro and Miguelote 1998; Sylos-Labini et al.\ 1998;
Pietronero and Sylos-Labini 2000).

The other side of the debate claims, however, that the traditional
statistical analysis of recent observations leads to the opposite
conclusion, {\it i.e.}, that the distribution of galaxies does homogenize
beyond a certain small scale (Peebles 1980, 1993; Davis 1997; Wu,
Lahav and Rees 1999). Besides, this group has a strong theoretical
rejection to fractals under the grounds that a possible fractal pattern
for the large-scale structure of the Universe cannot agree with what we
know about the structure and evolution of the Universe, as this knowledge
is based on the cosmological principle and the
Friedmann-Lema\^{\i}tre-Robertson-Walker (FLRW) spacetime, with both
predicting spatial homogeneity for the universal distribution of matter.
Therefore, this orthodox view also claims that such an observable
homogenization is necessary in order to ``make sense of a
Friedmann-Robertson-Walker universe'', since ``the FRW metric presumes
large scale homogeneity and isotropy for the Universe'', and ``in [this]
cosmological model, the mean density of the Universe is perfectly well
defined'' (Davis 1997; see also Wu, Lahav and Rees 1999). Moreover,
inasmuch as the cosmic microwave background radiation is isotropic, a
result predicted by the FLRW cosmology, this group is, understandably,
not prepared to, as it seems, give up the standard FLRW universe model
and the cosmological principle, as that would mean giving up most, if
not all, of what we learned about the structure and evolution of the
Universe since the dawn of cosmology (Peebles 1993; Davis 1997; Wu,
Lahav and Rees 1999; Mart\'{\i}nez 1999).

To reach those opposing conclusions, the validity of the methods used 
by both sides of this debate are, naturally, hotly disputed, and so far
there has not yet been achieved a consensus on this issue. However, even
if one is prone to part of the orthodox argument, {\it i.e.}, that we
cannot simply throw away some basic tenets of modern cosmology, like the
cosmological principle and the highly successful FLRW cosmological model,
when one looks in a dispassionate way at the impressive data presented
by the heterodox group, one cannot dispel a certain uneasy feeling that
something might be wrong in the standard observational cosmology: their
results are consistent and agree with one another (Coles 1998).

Actually, one thing in common between both sides of the debate is that
they seem to agree that if the distribution of galaxies does follow a
scaling, or fractal, pattern up to the limits of the presently available
observations, this would contradict the standard Friedmannian cosmology,
which in turn would lead to the dismissal of the cosmological principle
(Coleman and Pietronero 1992; Wu, Lahav and Rees 1999).\footnote{ \ The
heterodox group has recently retracted a bit from such more radical view,
arguing that an open Friedmann model may be compatible with a fractal
structure (Joyce et al.\ 2000). Some of their conclusions are similar to
the ones to be shown below, although the methods used to reach them are
completely different from the path taken here.}

It must be clearly noticed that the issues behind this debate are not
yet fractality in the sense of non-analytical sets, but so far it only
deals with properties of a smoothed-out and averaged fractal system,
whose main observational feature shows up as a decaying power-law type
behaviour for the average density as plotted against increasing
distances. So, when one talks about `observed fractality of the large-scale
distribution of galaxies', or this distribution as forming a `fractal
system', or, still, showing `fractal features', one means this type of
decaying average density power-law profile, which is consistent with
scale invariant structures typically featured in fractal sets
(Mandelbrot 1983; Pietronero 1987).

{From} this brief summary, it is clear that, to this date, the two sides
of The Fractal Debate seem to be locked in antagonistic and, as it may
initially appear, self-excluding viewpoints. Nevertheless, in order to
link these two positions to what is discussed in \S \ref{intro}, we
need first of all to start by carefully analysing the meaning behind
some terms used in this debate. The essential ones are `density' and
`observational density'. 

\subsection{Density Definitions in Cosmology}

The first aspect to note is that in cosmology we may define two types
of densities: a {\it local density} $\rho$ and an {\it average density}
$\langle \rho \rangle$, often also called {\it volume density} and
denoted by $\rho_v$. The latter is, of course, the local density
averaged over larger and larger distances. If the local density is
always the same, then local and average densities are equal, and, we
may suppose that in the standard spatially homogeneous cosmology we will always
have $\rho= \langle \rho \rangle$, as the local density is the same
everywhere. However, that would be a simplistic conclusion as in
standard cosmologies, both Newtonian and relativistic, the local density
is only uniform {\it at specific epochs}. In other words, it is a
function of time, $\rho=\rho(t)$, being the same everywhere only at
fixed epochs.

Suppose now two distances $d_1$ and $d_2$ such that $d_1 < d_2$. If
there exists a function $t=t(d)$ relating time and distance such that
bigger distances will mean earlier times, then an object located at
distance $d_1$ will be associated to time $t_1$, while another object
located at distance $d_2$ is associated to an earlier epoch $t_2$ $(t_1
> t_2)$. Since local density depends only on time, then $\rho(t_1) \not=
\rho(t_2)$, which means that the average $\langle \rho \rangle = (1/2)
[ \rho(t_1) + \rho(t_2)] \not= \rho(t_1)$ or $\rho(t_2)$. This
inequality between local and average density occurs {\it provided the
average is made along the curve $t=t(d)$}. Therefore, even in standard
cosmology local and average densities will only be equal at similar
epochs.

The point I wish to make here is that even in the standard cosmology
we can only talk about observing its spatial homogeneity if such
observational measurement is carried out at the same epoch $t$. In
other words, attempts to measure the smoothness of the Universe can
only make sense if one is assured to be doing so at similar epochs,
which, of course, must be within the observational errors of astronomical
observations. However, when observational cosmology deals with the
smoothness problem, this critical issue is usually vaguely dealt with,
by simply assuming that the sources should be close enough to be
guaranteed to be at the same, or very close, values of $t$. Usually the
implicit reasoning is that for ``small'' $z$ we should be roughly
observing at the same epochs, which are equal to our own (the observer),
while for ``large'' $z$ this is no longer the case. Such a reasoning
already implies that in standard cosmology there must be ranges where
$\langle \rho \rangle$ will start deviating from $\rho$, but one still
lacks a method for substantially narrowing the range from what one means
by ``large'' and ``small'' redshifts, allowing then a considerable grey
area between them. Not surprisingly, it is exactly in this grey area
that The Fractal Debate is thriving.

Anyway, the possible importance of the analysis above rests on
the existence of a function $t=t(d)$, and its possible relevance to
the smoothness problem in cosmology. As I shall show below, such a
function does exist when one uses Bondi's criterion discussed in
\S \ref{intro}, and is given by the past null geodesic. Since this
function is not defined in Newtonian cosmology, it can only be
obtained in a fully relativistic approach to cosmology, meaning that
one needs to use Bondi's criterion from the start. Then when one finds
the range where $\rho \not= \langle \rho \rangle$, that will
immediately give us the distance values where one has a breakdown of
the Newtonian approach to standard cosmology, at least as far as the
smoothness problem is concerned. Therefore, at the range where that
happens, the two simplifying hypotheses discussed in \S \ref{intro}
will no longer be applicable.

In hierarchical, or fractal, cosmologies, the problems above do not
appear as seriously as in the standard cosmology because it has been
realized long ago that in a fractal universe one needs to differentiate
local from average density from the start (Wertz 1970, 1971).
However, since these earlier models are only on Newtonian cosmology, it
was not possible to discuss their relativistic aspects, and how and
where they are related to standard cosmology. As I shall show below,
this is only possible when Bondi's criterion is applied to standard
cosmology {\it at the same time} as an average density is defined
in these models.

As final remarks, it is important to notice that the basic question
being dealt with above is how a spatially homogeneous cosmology may
appear, or be observed as, inhomogeneous, and the key to answering
it lies on how one constructs an
appropriate average. As shown above, this is done by averaging local
densities along the past null cone. Related to this question is the
opposite problem, which appears from time to time in the literature, and
reads as follows. Could a spatially inhomogeneous cosmological model
evolve on average like a spatially homogeneous universe model? This
problem is known as 
``the averaging problem in cosmology'', and its main difficulty is the
notion of average, whose specification and unambiguous definition is
not easy to establish, mainly because it is not straightforward to
produce an unique meaning to the averaging of tensors (Buchert 1997ab,
2000; Ellis 2000). The problem being discussed in this paper in a
sense belongs to the averaging problem in cosmology, but whose
motivation is opposite to the original question that established it.

\section{Local Versus Average Relativistic Density}\label{densities}

As we saw above the relationship between local and average cosmological
densities in standard cosmologies is not as straightforward as it may
initially appear, and implies some ambiguities and subtleties. The next
question then is how those subtleties will express themselves in a
relativistic setting. On this respect, Bonnor (1972) was the first to
point out that in a relativistic framework the local density is the
quantity entering into the energy-momentum tensor of Einstein's
equations, while the average density is obtained by averaging over a
sphere of a given volume, which may be arbitrarily large. He then
wrote an expression for the volume density, defined as the ratio
between the integrated mass $\int_0^r 4\pi \rho(\bar{r})
{\bar{r}}^2 d\bar{r}$, and a volume defined as $4 \pi r^3/3$. Here
$r$ is the radius coordinate (Bonnor 1972, eq.\ 3.4). Bonnor's
assumptions were a consequence of the impossibility of solving 
analytically the null geodesic equation of his model. Nevertheless, they
bring several conceptual problems for calculating observational averages
in cosmology, and unless we discuss them in detail their unchecked use
may lead to an unrealistic model.

In the first place, in order to relate the average density with
observations, it is necessary to integrate the local density
{\it along the past light}, which is the geometrical locus of astronomical
observations. This means solving the null geodesic equation, a task often
more difficult than solving Einstein's field equations themselves. In
addition, the best comparison with observations are obtained using number
counting rather than integrated mass (see below), but, to do so, we
also need the past null geodesic's solution. Secondly, taking the
radius coordinate as distance indicator is inconsistent
with Bondi's criterion outlined above. In general relativity coordinates
are labels to spacetime events, and, therefore, cannot be used as
distance, unless we are assured to be in a region of flat and Euclidean
geometry. However, to be assured of that we need to have a prior
method for determining up to where it is safe to use such
approximations, which is the object of the present discussion. Since
this is the aim of the method outlined here, we, therefore, cannot start
with this assumption and coordinates cannot be used as distances. Finally,
Bonnor's definition of volume density is not along the null cone, but at
hypersurfaces of constant time, where, astronomical observations
are not located.

Clearly, what we need is some volume definition that can be compared with
observational data. In other words, we need to define volume along the
null cone. But, to define a volume we need to choose some distance, and
since there are several ones in cosmology, we should first discuss a
method for choosing the appropriate distance definition required for
analysing the smoothness problem in cosmology.

\subsection{Cosmological Distances}\label{distances}

Distances in relativistic cosmology are a confusing subject, especially
because there is no unifying terminology. Nevertheless, the number of 
distance definitions are in fact quite small. As yet, the best treatment
of this issue has been given by Ellis (1971, p.\ 144), where one can find
a rigorous relativistic treatment of observations in cosmology, leading
to the most important cosmological distance definitions at section 6.4.3
of his paper. Moreover, Ellis' treatment is general in the sense that it is
valid for {\it any} cosmological model. Therefore, his analysis is not
restricted to the standard cosmology, as it is the case in virtually all
other treatments of this subject.

At a first thought one may think that the difficulties associated with
the distance concept in cosmology can be skipped if we were to use only
{\it separations}, also generically known as {\it proper distances}
(Ellis and Rothman 1993), that is, line element integrals
$\int ds$. The problem with such definitions comes from the fact
that separations are in general
unobservable geometrical distances. One example is the {\it absolute
distance} (d'Inverno 1992, p.\ 325), also known as {\it interval
distance} (Sandage 1995, p.\ 13), which would require us to place a
rigid rod between two astronomically separated points at the same epoch,
that is, assuming $dt=0$, as absolute distances are defined at constant
time hypersurfaces. This is not only observationally unfeasible, but
would also go against Bondi's criterion outlined in \S \ref{intro}.
Therefore, the absolute distance seems to be a device possessing little,
if any, relevance in discussing the observational smoothness of the
Universe. {\it Comoving distances} (also called {\it comoving coordinate
distances}) do not seem to be of much use in here too, since they are in fact
coordinate distances, or simply labels to spacetime events. In spherically
symmetric models the one most used is the comoving radial distance
coordinate, which, of course, requires the condition $dt=0$ in its
definition.\footnote{ \ It is common to call by {\it geodesic distance} the
separation between two points located over some hypersurface, or lower
dimensional surface. So the absolute distance is the geodesic distance
defined over the surface where $dt=d \theta= d\phi =0$ in spherically
symmetric metrics (d'Inverno 1992, p.\ 325). Longair (1995, p.\ 362) also
uses a similar terminology.}

Thus, we must turn our attention to discussing cosmological distances
along the backward null cone. However, along this hypersurface the
4-dimensional interval between two points is zero, that is, $ds=0$, and,
therefore, we must perform line element integrations over this specific
surface, given by $\int_c d \sigma$, where $C$ represents the null cone
hypersurface, or a line over it, and $d \sigma$ is the line element over
$C$, and is necessarily of lower dimension. The difficulty with this
procedure is that distances defined that way are not unique, a fact
which leaves us no choice but, to deal only with distances which can be
operationally defined by means of relationships among observational
quantities calculated along the null cone. The one mostly used in
astronomy is the {\it luminosity distance} $d_\ell$, defined as a
relationship between the observed flux $F$ of an astronomical source
and its intrinsic luminosity $L$, in a flat and static space. One can
also define the {\it observer area distance} $d_A$, also known simply
as {\it area distance}, and the {\it galaxy area distance} $d_G$. Both
$d_A$ and $d_G$ determine distances by comparing a solid angle measured
either at the observer or at some galaxy, and the intrinsic area of an
object. Since $d_G$ implies a knowledge of this solid angle {\it at the
galaxy}, it is unobservable, by definition (Ellis 1971). One can also
define a {\it parallax distance}, $d_p$, obtained by means of parallax
measurements (McCrea 1935, p.\ 296; Ellis 1971). Inasmuch as, under
our present astronomical technology, data on galaxy parallaxes are not
yet available, the parallax distance will play no role in this
discussion.\footnote{ \ Cosmological distances appear under different
names in the literature. For instance, the observer area distance
$d_A$ is called {\it angular diameter distance} by Weinberg (1972),
and {\it corrected luminosity distance} by Kristian and Sachs (1966).
The galaxy area distance $d_G$ is named {\it effective distance} by
Longair (1995, pp.\ 375), {\it angular size distance} by Peebles
(1993, pp.\ 319 and 328), and {\it transverse comoving distance}
by Hogg (1999). Such a profusion of names can only bring even more
confusion to the subject, specially as some of these names are
very unprecise from a geometrical viewpoint. In this paper I shall
follow Ellis' (1971) terminology as I believe his name choices are the
least confusing, and, geometrically, most appropriate.} As a final
remark, although $d_\ell$, $d_A$, $d_G$, and $d_p$ are the only
observationally accessible distances, it is conceivable that 
the observational distances could be related to distances which are not
directly observational. In this case one will require more use of
theory in order to provide such a link, but even so those unobservable
distances will always require the use of more directly observable
quantities. 

The observational distances above are linked by a remarkable geometrical theorem,
due to Etherington (1933), known as {\it reciprocity theorem} (see
also Ellis 1971; Schneider, Ehlers, and Falco 1992), which may
be written as follows,
\be
    d_\ell=d_A {\left( 1+z \right) }^2=d_G \left( 1+z \right).
    \lb{1}
\ee
In terms of bolometric (all wavelengths) flux-luminosity equations,
the reciprocity theorem yields,
\be
   F=\frac{L}{4 \pi { \left( d_A \right) }^2 {\left( 1+z \right) }^4}
    =\frac{L}{4 \pi { \left( d_G \right) }^2 {\left( 1+z \right) }^2}
    =\frac{L}{4 \pi { \left( d_\ell \right) }^2}.
   \lb{2}
\ee

Since all distances above have clear definitions, we may ask, which
one is correct? On this respect it is worth reproducing Allan Sandage's
wise remarks on this issue. ``We cannot measure distances by placing
rigid rods end to end. Rather, operational definitions of distance `by
angular size', `by apparent luminosity', `by light travel time', or `by
redshift' are perforce employed. Their use then requires a theory that
connects the observables (luminosity, redshift, angular size) with the
various notions of distances (McVittie 1974). One of the great initial
surprises is that these distances differ from one another at large
redshift, yet all have clear operational definitions. Which distance is
`correct?' {\it All} are correct, of course, each consistent with their
definition. Clearly, then, distance is a construct (...) operationally
defined entirely by its method of measurement.'' (Sandage 1988, p.\ 567).

In addition, he also offered a practical prescription regarding how to
deal with distances. ``The best that astronomers can do is to connect 
the observables by a theory and test predictions of that theory when
the equations are written in terms of the observables alone. To this end,
the concept of distance becomes of heuristic value only. It is simply
an auxiliary {\it parameter} that must drop from the final predictive
equations.'' (Sandage 1988, p.\ 567).

There is no question about the correctness of Sandage's remarks, and
the wisdom of his prescription. However, the question remains as to
what extent his prescription can be followed when dealing with the
possible observational homogeneity of the Universe. The reason for that
comes from the realization that the smoothness problem revolves around
the concept of observational mean density, which requires some definition
of volume, which, in turn, also requires some definition of distance.
Thus, when we deal with an observational mean density, the distance used
in its definition cannot be dropped from the final equations, and, so,
it is no longer an internal parameter, but a quantity which defines the
mean density itself. We therefore have to face the fact that {\it the
smoothness problem of the Universe requires us to choose a definition
of distance}. Sandage's prescription above is not applicable to this
problem. So, there is here some subjectivity in the sense that any
analysis of this issue will depend on the distance choice. If we do not
make this choice explicitly, it will enter implicitly in our problem,
by the back door.

At this stage one may be tempted to say that if we use the redshift
instead of any distance definition, these problems are solved. However,
the redshift is a distance indicator, and it will correspond to some
distance in some ranges. In fact, in Einstein-de Sitter cosmology it
scales with the luminosity distance at low $z$, and follows its 
asymptotic behaviour at the big bang (see below).

The reciprocity theorem gives us a general relation among cosmological
distances, but it does not tell us how to calculate them. {From}
equations (\ref{2}) it is clear that unless we have, in advance,
astrophysical information about intrinsic properties of the sources, the
only way we can solve equations (\ref{2}) is by assuming some cosmological
model, obtaining expressions for some previously chosen distance in the
assumed model, and feeding those expressions, together with observational
data, into equations (\ref{2}). However, in discussing the possible
observational smoothness of the Universe, we have an additional problem
to worry about. As we saw above, testing the observable galaxy
distribution homogeneity implies an implicit choice of distance. For
instance, if we collect data on apparent magnitude ($F$, in fact) and
does not make redshift corrections, we will end up with the luminosity
distance $d_\ell$, as it assumes a static and non-expanding universe.
On the other hand, if we do make redshift corrections, depending on the
used power of $(1+z)$ factors we may get either $d_G$ or $d_A$. In
principle, $d_A$ could be determined independently from equation
(\ref{2}), if, by some astrophysical consideration, we are able to infer
the intrinsic size of an object (Ellis 1971, p.\ 153). In practice,
however, this is a difficult task to be performed, and can only be done
to a small number of nearby objects. Due to this, such a knowledge
will not affect much our discussion here as the smoothness problem
requires us to know $d_A$ in large numbers, at the scale of present day
redshift surveys, which count thousands of galaxies. Thus, it should be
clear by now that, besides choosing a cosmological model, {\it the way
we collect and organize our astronomical data may be all that matters
in our implicit choice of distance}.

\subsection{Distances, Volumes and Densities}\label{both}

In a remarkable paper, McVittie (1974) showed that all observational
distances differ at large $z$, but are almost the same at moderate to
small redshifts. Based on this study, and at a quick look at equations
(\ref{1}), one may wonder if all previous discussion is irrelevant,
as all distance definitions should produce similar, if not equal,
results for $z<1$.

Here, however, another subtlety of the smoothness problem in cosmology
comes into play. While all distances are similar at small $z$, {\it the
observable homogeneity of the Universe is not discussed in terms of
distances, but in terms of average densities}. These are theoretically
constructed as being a ratio between number counting and observable
volumes, where the latter are themselves formed by third power of
distances, all that along the past null cone. Observable distances are,
however, non-linear functions of a null geodesic (unobservable) affine
parameter, meaning that {\it average densities are highly non-linear
functions along null geodesics}. So, a change in distance definition
can dramatically alter the behaviour of average density, no matter if
those distances are similar to each other at close values of the
affine parameter. We are dealing here with highly non-linear functions
along the past null cone, which means that simplistic predictions about
their behaviour can be very deceptive. We will return to this point
below, with specific examples. 

\subsubsection{An Example: the Einstein-de Sitter Cosmology}

Let us write the metric for the Einstein-de Sitter (EdS) model, the
simplest standard cosmology, as follows ($c=G=1$),
\be
  ds^2=dt^2-a^2(t) \left[dr^2+r^2 \left( d\theta^2 + \sin^2 \theta d
  \phi^2 \right) \right],
  \lb{metric}
\ee
where $a(t)$ is the scale factor, given by
\be
   { \left( \frac{da}{dt} \right) }^2= \frac{8 \pi}{3} \rho a^2(t),
   \lb{a}
\ee
and the local density is
\be
   \rho=\frac{1}{6\pi a^3(t)}.
   \lb{rho}
\ee
If we label our present time hypersurface ``now'' as $t=0$, then the
solution of equation (\ref{a}) may be written as
\be a(t) = { \left( t + \frac{2}{3H_0} \right) }^{2/3},
    \lb{sola}
\ee
where $H_0$ is today's value of the Hubble constant.

We can obtain the equation for the past light cone by integrating the past
null geodesic of metric (\ref{metric}), $dt/dr=-a(t)$, from ``here and now''
$(t=r=0)$ up to $t(r)$. The solution is given by
\be 3 { \left( t+ \frac{2}{3H_0} \right) }^{1/3}= { \left( \frac{18}{H_0}
    \right) }^{1/3} -r,
    \lb{null}
\ee
where the radius coordinate $r$ plays the role of the parameter along
the null geodesic.

As it is well known, the redshift in this cosmology is given by
$1+z= a(0)/a(t)$, or, using equation (\ref{sola}),
\be 1+z= { \left( \frac{18}{H_0} \right) }^{2/3}
         { \left[ { \left( \frac{18}{H_0} \right) }^{1/3} -r \right] }^{-2},
	 \lb{z}
\ee
since along the null cone the scale factor becomes,
\be a[t(r)]= \frac{1}{9} { \left[ { \left( \frac{18}{H_0} \right)
             }^{1/3} -r \right] }^2.
    \lb{a2}
\ee

The distances defined by equation (\ref{1}), may, in this cosmology,
be obtained by means of the area distance $d_A=r \; a[t(r)]$, (Ribeiro
1992b, 1995). Therefore, they may be written as,
\begin{eqnarray}
  d_A    & = & \frac{r}{9} { \left[ { \left( \frac{18}{H_0} \right)
               }^{1/3} -r \right] }^2, \lb{da} \\
  d_\ell & = & \frac{r}{9} { \left( \frac{18}{H_0} \right) }^{4/3}
               { \left[ { \left( \frac{18}{H_0} \right) }^{1/3} -r
	       \right] }^{-2}, \lb{dl} \\
  d_G    & = & r { \left( \frac{2}{3H_0} \right) }^{2/3}. \lb{dg}
\end{eqnarray}

The big bang singularity hypersurface is reached when metric
(\ref{metric}) degenerates at some early epoch. Let us call the big bang
time coordinate by $t_b$. This means that $a(t_b)=0$, and, with this
result we can also obtain the value of the null geodesic parameter
($r$, in this case) when the past light cone reaches the big bang. Doing
so, the big bag coordinates along the null geodesic may be written as,
\be t_b=- \frac{2}{3H_0}, \; \; \; r_b= { \left( \frac{18}{H_0} \right)
    }^{1/3}.
    \lb{bang}
\ee 

With these coordinates we can obtain the asymptotic behaviour for the
redshift and the three distances above as one approaches the big bang.
Thus, the following important limits hold in EdS cosmology,
\be \lim_{r \rightarrow r_b} z      = \infty, \; \; \; \; \; \; \;
    \lim_{r \rightarrow r_b} d_\ell = \infty, \; \; \; \; \; \; \;
    \lim_{r \rightarrow r_b} d_A    = 0,      \; \; \; \; \; \; \;
    \lim_{r \rightarrow r_b} d_G    = 2/H_0.
    \lb{lims}
\ee
Notice the completely different asymptotic behaviour of the distances.
The luminosity distance grows without bound, as well as the
redshift, while the galaxy area distance grows up to a maximum and
finite value. On the other hand, the area distance starts growing and
then decreases, reaching zero at the big bang. It is not difficult to
show that $d_A$ reaches a maximum at $z=1.25$. These results are similar
to McVittie's (1974), although here we have reached them by means of a
fully analytical study, that produced exact solutions.

If we now invert equation (\ref{dl}), we may write the distances in
terms of the redshift, as follows,
\begin{eqnarray}
  d_A    & = & \frac{2}{H_0} \left[ \frac{1+z- \sqrt{1+z}}{{ \left(
               1+z \right) }^2} \right]
	 =   \frac{z}{H_0}-\frac{7z^2}{4H_0}+ \frac{19z^3}{8H_0}-
               \ldots, \lb{daz} \\
  d_\ell & = & \frac{2}{H_0} \left( 1+z- \sqrt{1+z} \right)
         =   \frac{z}{H_0}+\frac{z^2}{4H_0}-\frac{z^3}{8H_0}+
               \ldots, \lb{dlz} \\
  d_G    & = & \frac{2}{H_0} \left( \frac{1+z- \sqrt{1+z}}{1+z}
               \right)
         =   \frac{z}{H_0}-\frac{3z^2}{4H_0}+\frac{5z^3}{8H_0}-
	       \ldots, \lb{dgz}
\end{eqnarray}
where the power series expansions hold for $z<1$. As expected, those
distances coincide on first order, but what we seek here is to determine
the influence on the average densities of higher order terms at
moderate redshift ranges.

Figure \ref{fig1} shows a plot of distances against redshift in EdS
cosmology. It is clear the different asymptotic behaviours, as well as the
deviation from one another at moderate values for $z$. Therefore, second
order terms play an important role at moderate redshift ranges, and
since average densities are built as third power of distances, one can
expect an even more important influence of second order terms on average
densities.
\begin{figure}[ht]
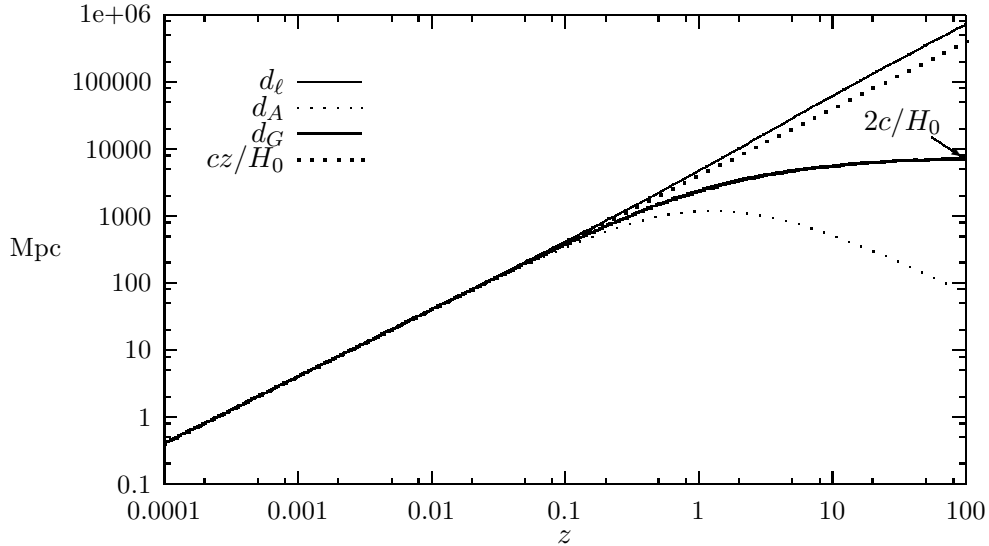

\begin{center}
\setlength{\unitlength}{0.240900pt}
\ifx\plotpoint\undefined\newsavebox{\plotpoint}\fi
\sbox{\plotpoint}{\rule[-0.200pt]{0.400pt}{0.400pt}}%

\caption{Plot of the different distance definitions against the redshift,
as given by equations (\protect\ref{daz}), (\protect\ref{dlz}), and
(\protect\ref{dgz}). The figure shows {\it exact solutions}. It is
clear the deviation at moderate redshift
ranges. The plot also shows the first order term for comparison, and one
may also notice that the luminosity distance provides the closest
scaling against this first order term, which is nothing more than the
Hubble law. Here $H_0=75$ km s$^{-1}$ Mpc$^{-1}$ and $c= 300,000$ km
s$^{-1}$.}
\lb{fig1}
\end{center}
\end{figure}

The next step on our analysis is to define the observational volume. It
is natural to use an extension of Euclidean volume, and when doing this
we end up with three different expressions for observational volume,
\be V_A=\frac{4}{3} \pi {d_A}^3, \; \; \; \; V_\ell=\frac{4}{3} \pi
    {d_\ell}^3, \; \; \; \;  V_G=\frac{4}{3} \pi {d_G}^3.
    \lb{vs}
\ee
We may also define a volume in the so-called ``redshift space'', as
follows,
\be V_z=\frac{4}{3} \pi {d_z}^3,
    \lb{vz}
\ee
where
\be d_z = \frac{c z}{H_0}.
    \lb{dz}
\ee
In the equation above the two constants are necessary for correct dimension,
and light velocity was included explicitly for clarity.

An interesting expression can be obtained from the equations above. If we
substitute the galaxy area distance, as given by equation (\ref{dgz}),
into the volume defined by this distance in equation (\ref{vs}), we
can easily obtain the following equation,
\be V_G=\frac{32 \pi}{3 {H_0}^3 { \left( 1+z \right) }^{3/2}} 
    { \left[ \sqrt{1+z} -1 \right] }^3.
    \label{volsand}
\ee
This volume definition is exactly the same as presented by Sandage (1995,
p.\ 19, eq.\ 1.42). The important point is that the expression above
appears in many standard texts as if it were the only possible expression
of volume as function of $z$, ignoring the fact that there are three other
volume definitions along the null cone which can be possibly used in any 
analysis, namely $V_A$, $V_\ell$, and $V_z$. So, while Sandage's (1995)
presentation of cosmological observational relations is correct, it is
not complete. One can define other types of observational volume and
density, whose definitions are relevant to the smoothness problem in
cosmology, and The Fractal Debate. In fact, it is the observational
importance of Etherington's reciprocity theorem that is currently being
under appreciated in observational cosmology. We must bear this point in
mind when discussing how other authors interpret cosmological observational
data (see below).  

To build expressions for the average density, the best method is to
calculate source number counting along the past light cone. {From} the
general expression derived by Ellis (1971, p.\ 159), we may obtain
the bolometric number counting in the EdS model (Ribeiro 1992ab), 
\be N_c=\frac{2 r^3}{9 M_g},
    \lb{nc}
\ee
where $M_g$ is the average galactic rest mass $(\sim 10^{11} M_\odot)$.

Now, average densities are easily calculated by means of the general
expression $\langle \rho \rangle=M_g Nc / V$. Since we have four types
of volume, we will end up with four different average densities, with
all being, in principle, obtainable from observational quantities.
Considering equations (\ref{z}), (\ref{daz}), (\ref{dlz}), (\ref{dgz}),
(\ref{vs}), (\ref{vz}), and (\ref{nc}), they may be written, as follows, 
\begin{eqnarray}
 \langle \rho_\ell \rangle & = & \frac{\rho_0}{{ \left( 1+z \right) }^3},
	 \lb{rl} \\
 \langle \rho_z \rangle & = &  8 \rho_0 { \left[ \frac{1+z-\sqrt{1+z}}{z \left(
	 1+z \right)} \right] }^3, \lb{rz} \\
 \langle \rho_A \rangle & = &  \rho_0 { \left( 1+z \right) }^3, \lb{ra} \\
 \langle \rho_G \rangle & = &  \rho_0,  \lb{rg}
\end{eqnarray}
where,
\be \rho_0= \frac{3 {H_0}^2} {8 \pi} \lb{r0} \ee
is the critical density, {\it i.e.}, the EdS local density at the present
time hypersurface.

The average density constructed with the galaxy area distance $d_G$,
given by  equation (\ref{rg}), behaves as today's local density, remaining
constant along the null cone. Therefore, if one uses such a distance in the
attempt to find some deviation from spatial homogeneity, even with data along
the null cone, one will find none simply because choosing $d_G$ leads to an
associated constant average density. This is a feature of EdS cosmology.
The other three averages are, therefore, the ones of importance for discussing
observational deviations from spatial homogeneity. Their behaviour are displayed
graphically in figure \ref{fig2}.
\begin{figure}[htb]
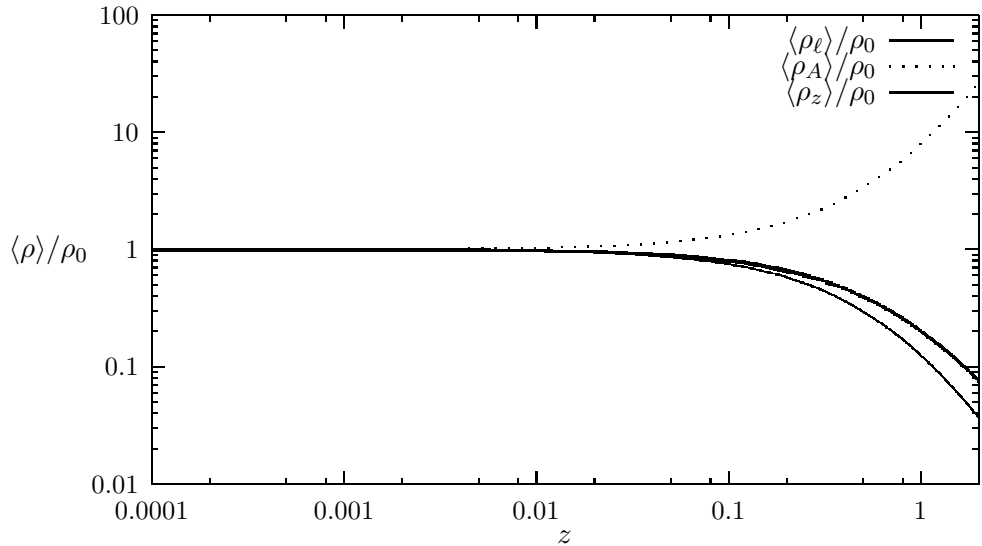

\begin{center}
\setlength{\unitlength}{0.240900pt}
\ifx\plotpoint\undefined\newsavebox{\plotpoint}\fi
\sbox{\plotpoint}{\rule[-0.200pt]{0.400pt}{0.400pt}}%

\caption{Plot of average densities $\langle \rho_\ell \rangle$ and
$\langle \rho_A \rangle$ respectively constructed using two different
distance definitions, $d_\ell$ and $d_A$. The third average density
$\langle \rho_z \rangle$, is constructed in ``redshift space'', that is,
using $z$ as distance in a volume definition given by equation
(\protect\ref{vz}). These average densities are plotted as a ratio
between them and the present time local density $\rho_0$. One can
clearly see that deviations from spatial homogeneity already occur at small
redshift ranges, becoming particularly large at $z \approx 0.1$.
Notice that at this same range the difference among the various
distances is still small, as shown in figure \protect\ref{fig1}. A
10\% deviation from $\rho_0$ occurs at $z \approx 0.04$ (see Ribeiro
1995 for detailed calculations of relativistic corrections at low
redshifts). Notice too the opposite behaviour of $\langle \rho_A
\rangle$ as compared to $\langle \rho_\ell \rangle$ and $\langle
\rho_z \rangle$.}
\lb{fig2}
\end{center}
\end{figure}

It is simple to see that when $z \rightarrow 0$, $\langle \rho_\ell
\rangle=\langle \rho_z \rangle=\langle \rho_A \rangle=\rho_0$.
So, these three averages tend to the present value of local density,
$\rho_0$, as they should. However, at the big bang, those averages will
behave very differently. As at the big bang $z \rightarrow \infty$
(eqs.\ \ref{lims}), one may show that the following limits hold,
\begin{eqnarray}
  \lim_{z \rightarrow \infty} \langle \rho_\ell \rangle & = & 0,
  \lb{lim3} \\
  \lim_{z \rightarrow \infty} \rho & = & \infty, \lb{lim1} \\
  \lim_{z \rightarrow \infty} \langle \rho_A \rangle & = & \infty,
  \lb{lim2} \\
  \lim_{z \rightarrow \infty} \langle \rho_z \rangle & = & 0. \lb{lim4}
\end{eqnarray}

These are remarkable results! Equation (\ref{lim3}) had already been
derived in Ribeiro (1992b), and equation (\ref{lim1}) is a well known
result in the literature. Equation (\ref{lim2}) is not that surprising,
since we know that the local density diverges at the big bang and we
would expect that an average density would do so as well. So, the big
surprise is the realization that {\it even in standard cosmological
models, some types of average densities vanish at the big bang}. This is
not restricted to EdS model, but occurs in all standard cosmologies
(Ribeiro 1993). The fact that there are vanishing average densities even
in standard cosmologies will be discussed below, in the context of The
Fractal Debate.

It is worth writing power series expansions of equations (\ref{rl}),
(\ref{rz}), and (\ref{ra}), as follows,
\begin{eqnarray}
 \langle \rho_\ell \rangle & = & \rho_0 \left( 1-3z+6z^2- 10z^3+ \ldots \right),
         \lb{prl} \\
 \langle \rho_z \rangle & = & \rho_0 \left( 1- \frac{9}{4}z +\frac{57}
         {16}z^2- \frac{39}{8}z^3+ \ldots \right),   \lb{prz} \\
 \langle \rho_A \rangle & = & \rho_0 \left( 1+3z+3z^2+z^3 \right).
         \lb{pra} 
\end{eqnarray}
Notice the existence of zeroth order terms in the expansions above,
while power series expansions for the distances, as given by equations
(\ref{daz}), (\ref{dlz}), and (\ref{dgz}) start on first order. As the
Hubble law is a distance-redshift law, derived from the first order
expansions of the distances, it is clear that due to the non-linearity
of the Einstein field equations, observational relations behave
differently at different redshift depths. Consequently, while the
linearity of the Hubble law is well preserved in the EdS model up to $z
\approx 1$ (Ribeiro 1995), a value implicitly assumed as the lower
limit up to where relativistic effects could be safely ignored, the
observational average densities constructed with $d_\ell$, $d_A$, and
$z$ are strongly affected by relativistic effects at much lower
redshift values. Then, while the zeroth order term vanishes in the
distance-redshift relation, it is non-zero for the average density as
plotted against redshift. This zeroth order term is the main factor
for the different behaviour of these two observational quantities at
small redshifts. Pietronero, Montuori and Sylos-Labini (1997) called
this effect as the ``Hubble-de Vaucouleurs paradox''. However, from
the analysis presented here, and in Ribeiro (1995; see also Abdalla,
Mohayaee and Ribeiro 2000), it is clear that this is not a paradox,
but just very different relativistic effects on the observables at
the moderate redshift range.

This effect explains why Sandage, Tammann and Hardy (1972) failed to
find deviations from uniform expansion in a hierarchical model: they
were expecting that such a strong observational inhomogeneity would
affect the velocity field, but it is clear now that if we take a
relativistic perspective for these effects they are not correlated at
the range expected by Sandage and collaborators. Notice that de
Vaucouleurs and Wertz also expected that their inhomogeneous hierarchical
models would also necessarily affect the velocity field, an effect also
conjectured by Pietronero (1987), and change the linearity of the Hubble law
at $z <1$, and once such a change was not observed by Sandage, Tammann,
and Hardy (1972) it was thought that this implied an immediate dismissal
of the hierarchical concept. Again, this is not necessarily the case
if we take a relativistic view of those observational quantities, as
prescribed by Bondi's criterion.

These findings can be summarized as follows. {\it The observational
inhomogeneity of EdS cosmology is not related to the linearity of the
Hubble law at moderate redshift ranges}. This conclusion can also be
extended to open and close standard cosmologies, but with some
limitations (Ribeiro 1993, 1995). Observers often use cosmological
formulae which does not follow Bondi's criterion, and so, they are
often under the assumption that at the scales where observations are
being made ($z < 1$) one can safely use the two simplifying assumptions
discussed in \S \ref{intro}, especially because in this range the Hubble
law is observationally verified to be very linear. However, we saw
above that  Hubble law linearity has a range which only coincides
with a constant density if we use the galaxy area distance $d_G$ as
distance definition. With all other averages that does not happen. Since
the observed average density is the key physical quantity for fractal
characterization (Pietronero 1987; Pietronero, Montuori and Sylos-Labini
1997; Coleman and Pietronero 1992; Sylos-Labini, Montuori and Pietronero
1998; Ribeiro and Miguelote 1998), we must seek hints for fractal
features in the behaviour of average densities which do not remain
constant along the null cone in EdS cosmology. 

Considering equations (\ref{lims}) we may rewrite equation (\ref{lim3}),
and conclude that in EdS cosmology the following limit holds,
\be
   \lim_{d_\ell \rightarrow \infty} \langle \rho_\ell \rangle=0.
   \lb {thelim}
\ee
This result provides a remarkable link to the hierarchical (fractal)
tradition. Thirty years ago James R.\ Wertz (1970, 1971) hypothesized
that a pure hierarchical cosmology ought to obey what he called {\it
``The Zero Global Density Postulate:} for a pure hierarchy the global
density exists and is zero everywhere'' (Wertz 1970, p.\ 18). Such a
result was also speculated by Pietronero (1987) as a natural
development of his fractal model. Therefore, what the above limit tells
us is that {\it the Einstein-de Sitter model does obey Wertz's zero
global density postulate, a key requirement of unlimited fractal
cosmologies}. This result appears naturally when one studies
cosmological observational relations in a fully relativistic setting.

In addition to the conclusion above, a quick look at figure \ref{fig2}
shows clearly that {\it two types of average densities decay at
increasing distances in EdS cosmology, this being another key aspect
of fractal cosmologies.} 

\subsubsection{Some Common Misconceptions}

In the light of the results above, we are now in position to discuss
some statements that appear in the literature about what the standard
and fractal cosmologies can, or cannot be. They are, in effect,
misconceptions, derived from no longer valid assumptions, as discussed
in \S \ref{intro}, which lead their authors to false problems. The
first important misconception is to state that finding homogeneity in
astronomical galaxy distribution data is the only way to make sense of
the FLRW cosmology, and not finding it leads to its falsification.

I showed above that some key fractal features appear in the EdS
cosmology. In addition, Ribeiro (1993, 1994) showed that they can also
be found in all standard cosmological models (see also Humphreys,
Matravers and Marteens 1998). These results were obtained without any
change in the model, its metric or its basic assumptions. So, those
observed fractal features appear {\it alongside} well-known features of
the model, like obedience of the cosmological principle, linearity of
Hubble law, CMBR isotropy. Moreover, cosmological parameters such as
$q_0$, $\Omega_0$, $H_0$ have their usual definitions and
interpretations. Therefore, recognizing observational fractality in
cosmology is not necessarily incompatible with well-known tenets of
modern cosmology. Nevertheless, the viewpoint currently being sustained
by both sides of The Fractal Debate is opposite to this one.

What is clear from all these results is that the homogeneity of the
standard cosmological models is {\it spatial}, that is, it is a {\it
geometrical} feature which does not necessarily translate itself into
an astronomically observable quantity (Ribeiro 1992b, 1993, 1994, 1995).
That happens only on special circumstances. Although a number of authors are
aware of this fact, what came as a surprise had been the calculated low
redshift value where this observational inhomogeneity appears (see
details in Ribeiro 1992b, 1995). Therefore, it is clear now that {\it
relativistic effects start to play an important role in observational
cosmology at much lower redshift values than previously assumed, at
least as far as the smoothness problem of the Universe is concerned.}

The second common misconception is to discuss the possible evidence
towards observational homogeneity/inhomogeneity in the Universe without
making explicit
the distance choice made in the analysis. To see how this difficulty
arises, let us try to clarify some puzzles surrounding The Fractal
Debate by asking the following question: which distance definition is being
implicitly used by the heterodox group? A thorough discussion of this
issue is beyond the scope of this paper, as it demands a detailed study
of the behaviour of these distances not on bolometric measurements, but
on limited frequency bandwidth, as this is how astronomical data is
gathered. The problem is that limited frequency range observational
relations alter the power of $(1+z)$ factors appearing in equations
(\ref{2}) (Ellis 1971; see also Ribeiro 1999), and we saw above how
dramatic such a change can be on the average densities. Other effects
must also be considered, like the luminosity function or K-correction,
which may alter even further the average densities, with unpredictable
results. Nevertheless, a sketchy discussion in bolometric terms can
be provided here.

If one takes redshift data and, by means of the Hubble law, transform
them into distances, by using the relation $cz=H_0d_z$, making no
further $(1+z)$ factors conversion, one
will be choosing as distance indicator the distance definition that scales
most closely with the Hubble law linearity. Figure \ref{fig1} showed that
this occurs with the luminosity distance. Figure \ref{fig2} showed that
an average density constructed that way decreases with higher distances.
Cappi {\it et al.\ }(1998) criticized Pietronero and collaborators
handling of data by not making the K-correction, which implies inclusion
of $(1+z)$ factors due to conversion from limited frequency bandwidth
observations to bolometric ones (Ribeiro 1999). That kind of conversion
can, therefore, destroy the fractal like decay of the average density, by
implicitly changing the distance definition. Therefore, {\it I suspect that
Pietronero and collaborators are systematically choosing $d_\ell$, or
$d_z$, as distance in their papers, while other authors may be using other
distances.} That may well explain the enormous difference in behaviour
that various authors, who are engaged in this debate, are finding
with the same data set.

To give another example of the difficulties generated when one ignores
the distance problem in cosmology, let us discuss the recent report
advanced by Pan and Coles (2000) where, by using a multifractal analysis
in the QDOT sample, they concluded that there is firm evidence towards
its observational homogenization at larger scales. Their study starts
by choosing a distance definition as given by Mattig's formula,
\be R=\frac{1}{H_0 {q_0}^2 (1+z)} \left[ q_0 z + \left( q_0 - 1 \right)
      \left( \sqrt{2q_0z+1} -1 \right) \right],
    \label{mattig}
\ee
where $R$ is their distance choice. In the context of this paper, the
obvious question is, what is $R$? In other words, which distance
definition are they implicitly choosing? They use the EdS model, and
then a trivial calculation taking $q_0=1/2$ reduces equation
(\ref{mattig}) to
\be R = \frac{2}{H_0} \left( \frac{1+z- \sqrt{1+z}}{1+z} \right).
    \label{mattig2}
\ee
Comparing with equation (\ref{dgz}) we conclude that
\be R=d_G. \label{mattig3} \ee
So, Pan and Coles have implicitly chosen the galaxy area distance to
carry out their analysis, which, then, continues by choosing cells of size
$R$ and then performing a multifractal measure. As seen above, $d_G$
is inappropriate for such kind of data analysis as it has the in-built
feature of showing no deviation from spatial homogeneity, even if the
Universe is of Friedmann type (see eq.\ \ref{rg}). The authors did not
provide any justification for the use of this equation, having in fact
ignored altogether the difficulties related to the distance choice
problem as discussed above. Their conclusions can, therefore, be
objected on the following grounds.

As they chose $d_G$, instead of $d_\ell$ or $d_A$, their results cannot be
related to the discussion made in here about fractal features in EdS
cosmology. In fact they cannot even be related to the data analysis
performed by Pietronero and collaborators as they have, most likely, been
using the luminosity distance. If Pan and Coles (2000) were to change
the chosen distance definition from galaxy area distance to luminosity
distance, or observer area distance, that would mean multiplying their
distance $R$ by a factor of $(1+z)$, or dividing by $(1+z)$, respectively.
Moreover, as they are using a flux limited sample, another $(1+z)$ factor
must be considered when changing from bolometric to flux limited measures
(Ellis 1971, p.\ 161). I wonder how those changes would modify their final
results. Consequently, their conclusions are of much narrower scope than
stated by the authors, and their analysis is inappropriate for probing
the possible observational inhomogeneity of the Universe.

\section{The Apparent Fractal Conjecture}\label{conjecture}

The results discussed above show that some key fractals features can
already be found in the simplest possible standard cosmological model,
that is, in the unperturbed Einstein-de Sitter universe. However, as
the average densities constructed with $d_\ell$ and $z$ do not decay
linearly in this model, considering all these aspects we may naturally
ask whether or not a perturbed model could turn the density decay at
increasing redshift depths into a power law type decay, as predicted
by the fractal description of galaxy clustering. If this happens, then
standard cosmology can be reconciled with a fractal galaxy distribution.
Notice that there are some indications that this is a real possibility,
as Amendola (2000) pointed out that locally the cold dark matter and
fractal models predict the same behaviour for the power spectrum, a
conclusion apparently shared by Cappi et al.\ (1998). In addition,
confirming Ribeiro's (1992b, 1995) conclusions, departures from the
expected Euclidean results at small redshifts were also reported by
Longair (1995, p.\ 398), and the starting point for his findings was
the same as employed in here and by Ribeiro (1992b, 1995): the use of
source number count expression along the null cone.

Considering all results outlined above, I feel there is enough grounds
to advance the following conjecture: {\it the observed fractality of the
large-scale distribution of galaxies should appear when observational
relations necessary for fractal characterization are calculated along
the past light cone in a perturbed metric of standard cosmology}. By
``observational relations necessary for fractal characterization'' I
mean choosing $d_\ell$ or $d_z$ as distances, and building average
densities with them, that is, deriving source number counting
expressions and calculating $\langle \rho_\ell \rangle$ and $\langle
\rho_z \rangle$ as defined above, all that along the past light cone.

This conjecture has theoretical and observational implications. On the
theoretical side, one can no longer ignore the distance choice, and all
calculations must clearly start with one. On the observational side, a
careful analysis is necessary about the way data is collected, reduced
and organized, as an implicit distance choice may occur during this
process.

If this conjecture proves, even partially, correct, fractals in
cosmology would no longer be necessarily seen as opposed to the
cosmological principle. Notice that this can only happen in circumstances
where fractality is characterized by an {\it observed}, smoothed out,
and averaged fractal system, as opposed to building a fractal structure
in the very spacetime geometrical structure, as initially thought
necessary to do for having fractals in cosmology (Mandelbrot 1983;
Ribeiro 1992a). Thus, the usual tools used in relativistic
cosmology, like the fluid approximation, will remain valid. As a
possible consequence of this conjecture, a detailed characterization
of the observed fractal structure could provide direct clues for the
kind of cosmological perturbation necessary in our cosmological models,
and this could shed more light in issues like galaxy formation.

A recent attempt to check the validity of this conjecture showed,
although in a restricted perturbative sense, that this conjecture is
sound as an apparent fractal pattern did emerge from the model
(Abdalla, Mohayaee, and Ribeiro 2000).

\section{Conclusions}\label{conclusion}

In this paper I have presented an analysis of the smoothness problem of
the Universe by focussing on the ambiguities arising from the
simplifying hypotheses aimed at observationally verifying whether or not
the large-scale distribution of galaxies is homogeneous. After briefly
reviewing The Fractal Debate, it is pointed out that in order to analyse
the possible observational homogeneity of the Universe one requires to make
a clear distinction between local and average density in a relativistic
framework. Then I showed that the different cosmological distance
definitions strongly affect the average density. An example, using the
Einstein-de Sitter cosmological model, is worked out, where I showed that
various observational average densities can be defined in this cosmology,
with the majority of them not leading to well defined values at
$z \approx 0.1$. I also revisited the discussion made in Ribeiro (1995),
which showed that the linearity of the Hubble law does not imply in an
observationally homogeneous density distribution of dust at the
moderate redshift ranges $(0.1 \le z < 1)$. Finally, I propose a
conjecture stating that the large-scale galaxy distribution should follow
a fractal pattern if observational relations necessary for fractal characterization are
evaluated along the past light cone. All these results were obtained
without any change in the standard cosmological models metric, meaning
that its observational fractality, as described in here, appears in cosmologies which
obey the cosmological principle, and have a near isotropy of the cosmic
microwave background radiation.

As discussed above, the divide caused by The Fractal Debate may not be
as radical as presented by both sides, and that it is possible to build
a bridge between both opinions, reconciling them by means of a change in
perspective regarding how we deal with observations in cosmology. What
was seen above is that there is already enough theoretical evidence to
suggest that the observed fractality can be accommodated {\it within} the standard
cosmology, where it would stem from the special way we are forced to
collect, organize and display our observational data on galaxy
distribution. And this special observational data collection and
organization are, in turn, a consequence of the underlining geometrical
structure of Friedmannian spacetime. Under this theoretical perspective,
the cosmological principle, uniform Hubble expansion, CMBR isotropy,
and well defined meanings for the cosmological parameters, such as
$\Omega_0$, can survive, together with the {\it observational
fractality} obtained by the heterodox group mentioned above. This
perspective has the advantage of preserving most of what we have learned
with the standard FLRW cosmology, and, at the same time, making sense
of Pietronero and collaborators' data, which, as seen above, can no
longer be easily dismissed.

At this point a relevant question arises immediately, and requires an
answer. If a FLRW cosmology may be observed to look like an universe
with properties of fractals, could this effect be nullified by
using relevant distances to match the way observations have actually
been carried out? In order to answer this question, it is important to
point out first of all that actual observations on galaxy distribution
basically consist of tables listing integrated $F$, $z$, and projected
angle positions on the sky. These are the real observations, and all
else is theory and interpretation. To state that there could be a manner
such that one can mimic the ways in which observations have ``actually'',
or ``really'' been carried out is the same as to say that there is an
actual, or real, cosmological distance. As we have seen above, 
searching for a ``true'' distance is a futile exercise, and
the same is also true regarding attempts to say how observations are
actually made. It is the handling of data that matters here, and not
the actual observations.

Therefore, at this stage it must be very clear that the essential
issues dealt with in this paper can be summarized as follows.
While observers have been handling astronomical data for quite a
while, and claiming that they are consistent with a spatially
homogeneous FLRW cosmology, the main point raised in here is that,
from a theoretical viewpoint, the same FLRW cosmology may be consistent
with an universe with observed fractal properties, as supported by data collected
by Pietronero and collaborators. In fact, claims that the only way to
make sense of FLRW model is by observing homogeneity, ignores the
richness of the standard model by sticking to a somewhat narrow
interpretation of its observables features. The various ways that data
is handled (e.g., K-correction) and effects are considered (e.g.,
galaxy evolution) will affect interpretation of data, and what I
have discussed above is that fractal properties might also be part
of those interpretations, and should not be considered as extraneous,
irrelevant, or wrong cosmological data handling. Indeed, if the
observational fractality
is one of the many possible interpretations of galaxy distribution data,
one may speculate that its fractal dimension may become an important
cosmological parameter, perhaps to be taken into consideration in any
model of galaxy formation.

Another way of summarizing the results of this paper is by noticing that
while observational cosmologists were aware of of the possible apparent
inhomogeneity of the standard model, which may also be called densities
changing with time, or, still, lookback effects, it is clear that this
phenomenon occurs at close ranges in EdS cosmology. The link with
fractal properties, that is, a smoothed-out and averaged fractal system
possessing properties of power-law average density decay, as originally
proposed by Pietronero (1987), but whose roots can be found in Wertz's
(1970) work, occurs because some of these observational densities decay,
rather than increase, at deeper distances, or, which is the same, earlier
times. Cosmologists have been working with the hypotheses that, (1) this
effect should not be important for $z<1$, {\it i.e.}, the effect of densities
changing with time is not relevant for the observational determination
of whether or not the Universe is observationally homogeneous, and (2),
because the local density diverges at the big bang, the same should
happen to all density definitions. It should be clear by now that 
these two hypotheses have difficulties when we consider the full
consequences of the reciprocity theorem in observational quantities.
Therefore, the analysis presented in standard texts is not complete.
There is more to be said on those issues than can be found in standard
texts, and this paper attempts at adding some ideas and results in
the context of the possible observational homogeneity of the Universe.

Reconciling two seemingly disparate universe models through the
recognition of the importance of a previously dismissed physical effect
would not be new in observational cosmology. During the 1910's there was
observational evidence supporting two opposing models about the size and
structure of the Milky Way, the {\it Kapteyn Universe} and {\it Shapley's
Model}. The former sustained that the Sun was located near the centre of
an approximately oblate spheroidal distribution of stars, whose dimension
was estimated to be about 8.5kpc, while the latter located the Sun at
the edge of the stellar system and estimated the size of the Milky Way
as being 100kpc. As at that time the nature of the nebulae, that is,
the question of whether or not they were structures belonging to the Milky
Way or external objects placed at distances much greater than the size of
the Milky Way, was still an unresolved issue (solved later by Hubble),
these two models were effectively dealing with the observable Universe of
the time, and, therefore, such a discrepancy was, perhaps, the main puzzle
in observational cosmology of that epoch, and which effectively led to a split
in the astronomical community. The public confrontation of these two views
took place in April 1920, and this event is now known as ``The Great
Debate'', although the final reconciliation between them was only reached
in the 1930's, when astronomers generally recognized that the apparent
stellar distribution is dominated by the effects of absorption. Kapteyn
himself allowed this possibility, but as he only considered Rayleigh
scattering as the possible source of obscuration, he dismissed this effect
once he found it to be small, while we now know that the dominant obscuration
source is dust absorption. On perspective, it is clear now that both sides
of the debate had elements of the truth, as we perceive it today, and even
the heliocentric Kapteyn Universe is not that absurd since the Sun does lie
close to the centre of a local loose cluster of stars.\footnote{\ See Binney
and Merrifield (1998, pp.\ 5-15), for a historical overview of this earlier
split of the astronomical community.}

The historical lesson to be learned from this episode is that controversial
issues in cosmology are not necessarily solved with the simple dismissal
of one side of the debate, as happened to be the case in the 1920's
controversy of static versus expanding universe or, later, the steady state
versus evolving universe, in the 1960's. Based on the theses exposed above,
it is the opinion of this author that The Fractal Debate may well be
overcome in a similar manner as the issues surrounding The Great Debate,
but in this case only when one recognizes that relativistic effects and
their consequences must be {\it fully} considered in observational cosmology.

\begin{flushleft}
{\Large \bf Acknowledgements}
\end{flushleft}
The origins of many ideas appearing in this paper can be traced back to
a series of very fruitful discussions I had with Malcolm A.\ H.\
MacCallum about a decade ago. At the time he was also the first to
suggest that the heterodox group should be taking the luminosity
distance as their distance choice. I am grateful to him for
that interesting and rich exchange of ideas. I am also grateful to W.\ R.\
Stoeger for discussions on relativistic density definitions while
visiting the University of Arizona five years ago. Finally, I thank two
referees for helpful and useful remarks, which improved the text.
Partial support from FUJB-UFRJ is also acknowledged.
\begin{flushleft}
{\Large \bf References}
\end{flushleft}
\begin{description}
\item Abdalla, E., Mohayaee, R., and Ribeiro, M.\ B.\ 2000, astro-ph/9910003
\item Amendola, L.\ 2000, {\it Proc.\ of the IX Brazilian School of Cosmology
      and Gravitation}, M.\ Novello, in press
\item Amoroso Costa, M.\ 1929, {\it Annals of Brazilian Acad.\ Sci.}, 1, 51,
      (in Portuguese)
\item Binney, J., and Merrifield, M. 1998, {\it Galactic Astronomy},
      (Princeton University Press)
\item Bondi, H., 1960, {\it Cosmology}, 2nd ed., (Cambridge University Press)
\item Bonnor, W.\ B.\ 1972, \mnras, 159, 261
\item Buchert, T.\ 1997a, Proc.\ 2nd SBF Workshop on Astro-Particle
      Physics, Ringberg 1996, proc.\ series SBF375/P002, R.\ Bender et al., 71,
      astro-ph/9706214
\item Buchert, T.\ 1997b, \aea, 320, 1, astro-ph/9510056
\item Buchert, T.\ 2000, 9th JGRG Meeting, Hiroshima 1999, invited
      paper, gr-qc/0001056
\item Callan, C., Dicke, R.\ H., and Peebles, P.\ J.\ E.\ 1965, {\it
      American J.\ Phys.}, 33, 105
\item Cappi, A., Benoist, C., da Costa, L.\ N., and Maurogordato, S.\ 1998,
      \aea, 335, 779
\item Charlier, C.\ V.\ L.\ 1908, {\it Ark.\ Mat.\ Astron.\ Fys.}, 4, 1
\item Charlier, C.\ V.\ L.\ 1922, {\it Ark.\ Mat.\ Astron.\ Fys.}, 16, 1
\item Coleman, P.\ H., and Pietronero, L.\ 1992, \physrep, 213, 311
\item Coles, P.\ 1998, \nat, 391, 120
\item Davis, M.\ 1997, {\it Critical Dialogues in Cosmology}, N.\ Turok,
      (Singapore: World Scientific), 13, astro-ph/9610149
\item de Vaucouleurs, G.\ 1970a, \sci, 167, 1203
\item de Vaucouleurs, G.\ 1970b, \sci, 168, 917
\item de Vaucouleurs, G., and Wertz, J.\ R.\ 1971, \nat, 231, 109 
\item d'Inverno, R.\ 1992, {\it Introducing Einstein's Relativity},
      (Oxford: Clarendon Press)
\item Disney, M.\ J.\ 2000, \grg, 32, 1125, astro-ph/0009020
\item Ellis, G.\ F.\ R.\ 1971, {\it General Relativity and Cosmology}, Proc.\
      of the International School of Physics ``Enrico Fermi'', R.\ K.\
      Sachs, (New York: Academic Press), 104
\item Ellis, G.\ F.\ R.\ 2000, \grg, 32, 1135
\item Ellis, G.\ F.\ R., and Rothman, T.\ 1993, {\it American J.\ Phys.},
      61, 883
\item Etherington, I.\ M.\ H.\ 1933, {\it Phil.\ Mag.}, 15, 761; reprinted,
      \grg, in press
\item Fournier D'Albe, E.\ E.\ 1907, {\it Two New Worlds: I The Infra World;
      II The Supra World}, (London: Longmans Green)
\item Harrison, E.\ R.\ 1993, \apj, 403, 28 
\item Harrison, E.\ R.\ 2000, {\it Cosmology}, 2nd ed., (Cambridge University Press)
\item Hogg, D.\ W.\ 1999, astro-ph/9905116
\item Humphreys, N.\ P., Matravers, D.\ R., and Marteens, R.\ 1998, {\it Class.\
      Quantum Grav.}, 15, 3041, gr-qc/9804025
\item Joyce, M., Anderson, P.\ W., Montuori, M., Pietronero, L., and Sylos-Labini,
      F.\ 2000, {\it Europhys.\ Lett.}, 49, 416, astro-ph/0002504
\item Keynes, J.\ M.\ 1936, {\it The General Theory of Employment, Interest,
      and Money}, preface
\item Kristian, J., and Sachs, R.\ K.\ 1966, \apj, 143, 379
\item Longair, M.\ S.\ 1995, {\it The Deep Universe}, Saas-Fee Advanced Course 23,
      B.\ Binggeli and R.\ Buser, (Berlin: Springer), 317
\item Mandelbrot, B.\ B.\ 1983, {\it The Fractal Geometry of Nature}, (New
      York: Freeman)
\item Mart\'{\i}nez, V.\ J.\ 1999, \sci, 284, 445
\item McCrea, W.\ H. 1935, {\it Zeit.\ f\"ur Astrophysik}, 9, 290
\item McCrea, W.\ H., and Milne, E.\ A.\ 1934, {\it Quart. J. Math. (Oxford
      Ser.)}, 5, 73; reprinted in 2000, \grg, 32, 1949 
\item McVittie, G.\ C.\ 1974, {\it Quart.\ J.\ Royal Astr.\ Soc.}, 15, 246
\item Milne, E.\ A.\ 1934, {\it Quart.\ J.\  Math.\ (Oxford Ser.)}, 5,
      64; reprinted in 2000, \grg, 32, 1939
\item Oldershaw, R.\ L.\ 1997, http://www.amherst.edu/$\sim \:$rlolders/LOCH.HTM
\item Pan, J., and Coles, P. 2000, \mnras, in press, astro-ph/0008240
\item Peebles, P.\ J.\ E.\ 1980, {\it The Large-Scale Structure of the
      Universe}, (Princeton University Press)
\item Peebles, P.\ J.\ E. 1993, {\it Principles of Physical Cosmology},
      (Princeton University Press)
\item Pietronero, L.\ 1987, {\it Physica} A, 144, 257
\item Pietronero, L., Montuori, M., and Sylos-Labini, F.\ 1997, {\it Critical
      Dialogues in Cosmology}, N.\ Turok, (Singapore: World Scientific),
      24, astro-ph/9611197
\item Pietronero, L., and Sylos-Labini, F.\ 2000, Proc.\ 7th Course in
      Astro-Fundamental Physics, Erice 1999, in press, astro-ph/0002124
\item Ribeiro, M.\ B.\ 1992a, \apj, 388, 1 
\item Ribeiro, M.\ B.\ 1992b, \apj, 395, 29 
\item Ribeiro, M.\ B.\ 1993, \apj, 415, 469 
\item Ribeiro, M.\ B.\ 1994, {\it Deterministic Chaos in General Relativity},
      D.\ Hobbil, A.\ Burd, and A.\ Coley, (New York: Plenum Press), 269
\item Ribeiro, M.\ B.\ 1995, \apj, 441, 477, astro-ph/9910145 
\item Ribeiro, M.\ B.\ 1999, gr-qc/9910014
\item Ribeiro, M.\ B.\ 2001, gr-qc/9909093
\item Ribeiro, M.\ B., and Miguelote, A.\ Y.\ 1998, {\it Brazilian J.\
      Phys.}, 28, 132, astro-ph/9803218
\item Ribeiro, M.\ B., and Videira, A.\ A.\ P.\ 1998, {\it Apeiron}, 5,
      227, physics/9806011
\item Sandage, A.\ 1988, \araa, 26, 561
\item Sandage, A.\ 1995, {\it The Deep Universe}, Saas-Fee Advanced Course 23,
      B.\ Binggeli and R.\ Buser, (Berlin: Springer), 1
\item Sandage, A., Tammann, G.\ A., and Hardy, E.\ 1972, \apj, 172, 253 
\item Sandage, A., and Tammann, G.\ A.\  1975, \apj, 196, 313 
\item Schneider, P., Ehlers, J., and Falco, E.\ E.\ 1992, {\it Gravitational
      Lenses}, (Berlin: Springer)
\item Sciama, D.\ W.\ 1993: {\it Modern Cosmology and the Dark Matter Problem},
      (Cambridge University Press)
\item Sylos-Labini, F., Montuori, M., and Pietronero, L.\ 1998, \physrep,
      293, 61,\\astro-ph/9711073
\item Weinberg, S.\ 1972, {\it Gravitation and Cosmology}, (New York: Wiley)
\item Wertz, J.\ R.\ 1970, {\it Newtonian Hierarchical Cosmology}, PhD thesis
      (University of Texas at Austin)
\item Wertz, J.\ R.\ 1971, \apj, 164, 227
\item Wu, K.\ K.\ S., Lahav, O., and Rees, M.\ J.\ 1999, \nat, 397, 225,
      astro-ph/9804062
\end{description}
\end{document}